\documentclass[prd,aps,showpacs,nofootinbib,superscriptaddress,11pt]{revtex4-1}
\usepackage{amsmath,amssymb}
\usepackage{epsfig}
\usepackage{color}
\usepackage{bm}
\usepackage{epstopdf}

\newcommand{\X}{X(3872)}
\newcommand{\DD}{D^0\bar{D}^0}
\newcommand{\Tr}{\text{Tr}}

\makeatletter
\def\slashchar#1{{\mathpalette\c@ncel{#1}}} 
\makeatother
\def\vsl{\slashchar{v}}

\begin{document}

\title{Detecting the long-distance structure of the $X(3872)$  
}

\author{Feng-Kun Guo}
\email{fkguo@hiskp.uni-bonn.de}
\affiliation{Helmholtz-Institut f\"ur Strahlen- und
             Kernphysik and Bethe Center for Theoretical Physics, \\
             Universit\"at Bonn,  D-53115 Bonn, Germany}
\author{Carlos Hidalgo-Duque}
\email{carloshd@ific.uv.es}
\affiliation{Instituto de F\'isica Corpuscular (IFIC),
             Centro Mixto CSIC-Universidad de Valencia,
             Institutos de Investigaci\'on de Paterna,
             Aptd. 22085, E-46071 Valencia, Spain}
\author{Juan Nieves}
\email{jmnieves@ific.uv.es}
\affiliation{Instituto de F\'isica Corpuscular (IFIC),
             Centro Mixto CSIC-Universidad de Valencia,
             Institutos de Investigaci\'on de Paterna,
             Aptd. 22085, E-46071 Valencia, Spain}
\author{Altug Ozpineci}
\email{ozpineci@metu.edu.tr}
\affiliation{Middle East Technical University - Department of Physics
TR-06531 Ankara, Turkey}
\author{Manuel Pav\'on Valderrama}
\email{pavonvalderrama@ipno.in2p3.fr}
\affiliation{Institut de Physique Nucl\'eaire,
             Universit\'e Paris-Sud, IN2P3/CNRS,
             F-91406 Orsay Cedex, France}

\begin{abstract}
\rule{0ex}{3ex} 
We study the $\X \to \DD\pi^0$ decay within a $D\bar D^*$ molecular picture for 
the $\X$ state. This decay mode is more sensitive to the long-distance   
structure of the $\X$ resonance than its  $J/\psi\pi\pi$ and $J/\psi3\pi$ 
decays, which are mainly controlled by the details of the $\X$ wave function at 
short distances.   We show that the $\DD$ final state interaction can be 
important, and that a precise measurement of this partial decay width can 
provide valuable information on the interaction strength between 
the $D^{(*)}\bar D^{(*)}$ charm mesons. 
\end{abstract}


\maketitle

\section{Introduction}
\label{sec:introduction}

It has been long since mesonic molecules in the charm sector were first 
theorized \cite{Voloshin:1976ap,De Rujula:1976qd} but there was not any 
experimental observation 
until the discovery of the $\X$ in 2003 in the $J/\psi \pi \pi$ channel 
\cite{Choi:2003ue}. However, important details of the inner structure of the 
resonance are still under debate. Among many different interpretations of 
the $\X$, the one assuming it to be\footnote{From now on, when we refer to $D^0 \bar D^{*0} ,
  D^+\bar D^{*-}$, or in general $D \bar D^*$ we are actually referring to the
combination of these states with their charge conjugate ones in order to
form a state with positive C-parity.} a $\left(D\bar
D^*-D^*\bar D \right)/\sqrt{2}$,  hadronic molecule 
(either a bound state~\cite{Tornqvist:2004qy} or a virtual 
state~\cite{Hanhart:2007yq}) with quantum 
numbers $J^{PC} = 1^{++}$ (as recently confirmed in Ref.~\cite{Aaij:2013zoa}) 
is the most promising.
Quite a lot of work has been done under this assumption, for reviews, see, for 
instance, Refs.~\cite{Swanson:2006st,Brambilla:2010cs}.
\begin{figure}[tb] 
\begin{center}
\includegraphics[width=0.6\textwidth]{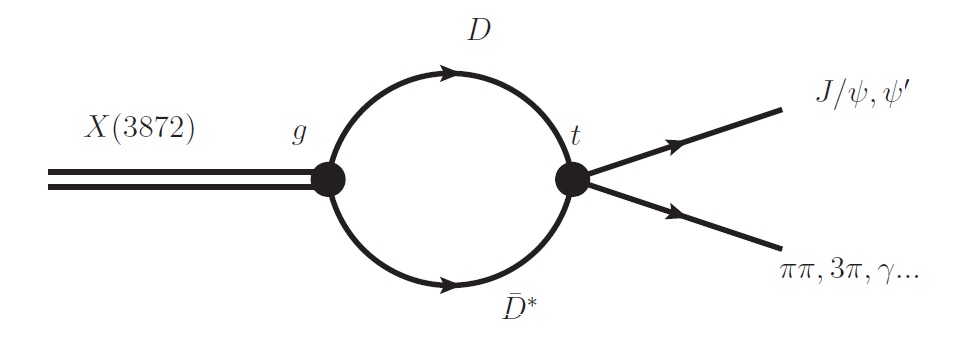}
\caption{Mechanism for the decays of the $X(3872)$ into $J/\psi\pi\pi$, 
$J/\psi3\pi$,
  $J/\psi\gamma$, $\psi'\gamma$ ...  assuming the $X(3872)$ to
be a $D\bar D^*$ molecule. The charge conjugated channel is 
not plotted.}\label{fig:shortdis}
\end{center}
\end{figure}
The most discussed decay channels of the $\X$ are those with a
charmonium in the final state, which include the $J/\psi\pi\pi$,
$J/\psi3\pi$, $J/\psi\gamma$ and $\psi'\gamma$. In the hadronic
molecular picture, these decays occur through the mechanism depicted
in Fig.~\ref{fig:shortdis}. Thus, the charm and anti-charm mesons
only appear in the intermediate (virtual) state, and the amplitude of
such decays is proportional to the appropriate charged or neutral $ D
\bar D^*$ loop integrals~\cite{Gamermann:2009fv}. Because the quarks in the two 
mesons have to recombine to get the charmonium in the final state, the 
transition from the charm--anti-charm meson pair into the $J/\psi$
plus pions (or a photon), occurs at a distance much smaller than both
the size of the $\X$ as a hadronic molecule ($\sim$ few fm's)\footnote{This is
  approximately given by $1/\sqrt{2\mu \epsilon_X} $ fm where $\mu$ is the
  reduced mass of the $D$ and $\bar D^*$ pair and $\epsilon_X =
  M_{D^0} + M_{D^{*0} } - M_{\X} =
  0.16\pm0.26~\text{MeV}$~\cite{Beringer:1900zz}.} and the range of
forces between the $D$ and $\bar D^*$ mesons which is of the order of 
$1/m_\pi\sim 1.5$ fm. In this case, if this transition matrix $t$ in
Fig.~\ref{fig:shortdis} does not introduce any momentum dependence,
the loop integral reduces to the wave function of the $\X$ at
the origin, $\Psi(\vec{0})$,~\footnote{The relative distance between the two 
mesons is zero in 
the wave function at the origin.} (more properly, around the origin, the 
needed ultraviolet regulator, for 
which we do not give details here, would smear the wave functions) 
\cite{Gamermann:2009uq}\footnote{For related discussions
  in case of the two-photon decay width of a loosely bound hadronic
  molecule, see Ref.~\cite{Hanhart:2007wa}.},
\begin{equation}
\langle f|\X \rangle = \int d^3 q \underbrace{\langle f | D \bar D^*(\vec{q}\,)\rangle}_{t}
\underbrace{\langle  D \bar D^*(\vec{q}\,)|\X\rangle}_{\Psi(\vec{q})} = t \int
  d^3 q \Psi(\vec{q}\,) = t\, \Psi(\vec{0}) = t\, {\hat g}G 
\label{eq:psi0}
\end{equation} 
where $\hat g$ is the coupling of the $\X$ to the $D \bar
D^*(\vec{q}\,)$ pair and $G$ is the diagonal loop function for the two
intermediate $D$ and $\bar D^*$ meson propagators, with the
appropriate normalizations that will be discussed below. The last
equality follows from the expression of the momentum space wave
function~\cite{Gamermann:2009uq}
\begin{equation}
\Psi(\vec{q}\,)=\frac{\hat g}{ M_X-M_D-M_{D^*}-\vec{q}^{\,2}/(2\mu) },
\end{equation}
where $\mu$ is the reduced mass of the $D$ and $\bar D^*$. This can be easily 
derived from the Schr\"odinger equation assuming the coupling of the $\X$ to 
$D\bar D^*$ to be a constant, which is valid since the $\X$ is very close to the 
threshold.
Thus, one can hardly extract information on the long-distance
structure of the $\X$ from these decays. 

In general, to be sensitive to the long-distance 
part of the wave function of a hadronic molecule, it is better to investigate 
the decay processes with one of the constituent hadrons in the 
final state and the rest of the final particles being products of the decay of 
the other constituent hadron of the molecule. 
For instance, for the case of the $\X$ as a $D\bar D^*$ molecule, we should 
use the $\X\to \DD\pi^0$ or $\X\to D\bar D\gamma$ to study the long-distance 
structure. In these processes, the relative
distance between the $D\bar D^*$ pair can be as large as
allowed by the size of the $\X$ resonance, since the final state is
produced by the decay of the $\bar D^*$ meson instead of a
rescattering transition. These decay modes have been addressed in some
detail in different works, e.g.,
Refs.~\cite{Swanson:2003tb,Voloshin:2003nt,Voloshin:2005rt,
  Fleming:2007rp, Liang:2009sp, Stapleton:2009ey, Baru:2011rs}. The $D^0\bar
D^0\pi^0$ mode has been already observed by the Belle
Collaboration~\cite{Gokhroo:2006bt,Adachi:2008sua}, which triggered
the virtual state interpretation of the $\X$~\cite{Hanhart:2007yq},
and it will be studied in detail in this work. 

On the other hand, heavy-quark spin symmetry (HQSS) and flavor symmetry
\cite{Isgur:1989vq,Isgur:1989ed, Neubert:1993mb, Manohar:2000dt} have
been widely used to predict partners of the $\X$
state~\cite{Guo:2009id,Bondar:2011ev,Voloshin:2011qa,Mehen:2011yh,
  Nieves:2011zz,Nieves:2012tt,HidalgoDuque:2012pq,Guo:2013sya,Guo:2013xga}.
Moreover, HQSS heavily constrains also the low-energy interactions
among heavy hadrons \cite{Bondar:2011ev, Mehen:2011yh, Nieves:2012tt,
  HidalgoDuque:2012pq, AlFiky:2005jd}. As long as the hadrons are not
too tightly bound, they will not probe the specific details of the
interaction binding them at short distances. Moreover, each of the
constituent heavy hadrons will be unable to see the internal structure
of the other heavy hadron. This separation of scales can be used to
formulate an effective field theory (EFT) description of hadronic
molecules~\cite{Mehen:2011yh,Nieves:2012tt,Baru:2013rta,Jansen:2013cba} compatible
with the approximate nature of HQSS.  At leading order (LO) the EFT is
particularly simple and it only involves energy-independent contact
range interactions, since pion exchanges and coupled-channel effects
can be considered
subleading~\cite{Nieves:2012tt,Valderrama:2012jv}. Moreover, the
influence of three-body $D\bar D\pi$ interactions on the properties of
the $\X$ was found to be moderate in a Faddeev
approach~\cite{Baru:2011rs}. In particular since we will only be
interested in the $\X$  mass and its couplings to the neutral and
charged $D\bar D^*$ pairs, the three-body cut can be safely neglected
at LO. 
As a result of the HQSS, assuming the $\X$ being a $D\bar D^*$ molecular
state, it is expected to have a spin 2 partner, a $D^*\bar D^*$
$S$-wave hadronic
molecule~\cite{HidalgoDuque:2012pq,Nieves:2012tt,Guo:2013sya}.  This
is because the LO interaction in these two systems are exactly the
same due to HQSS. The interaction between a $D$ and a $\bar D$, on the
contrary, is different. It depends on a different combination of low
energy constants
(LECs)~\cite{Mehen:2011yh,Nieves:2012tt,HidalgoDuque:2012pq}. The
$\X\to \DD\pi^0$ decay, on one hand, detects the long-distance
structure of the $\X$, on the other hand, it provides the possibility
to constrain the $D\bar D$ $S$-wave interaction at very low energies.
Hence, it is also the purpose of this paper to discuss the effect of
the $D\bar D$ $S$-wave final state interaction (FSI) in the 
$\X\to\DD\pi^0$ decay, which can be very large
because of the possible existence of a sub-threshold isoscalar
state in the vicinity of 3700 MeV~\cite{Gamermann:2006nm, 
Nieves:2012tt,HidalgoDuque:2012pq}.

As mentioned above, this $\X$ decay channel has been previously
studied. The first calculation was carried out in
Ref.~\cite{Voloshin:2003nt} using effective-range theory. In 
Ref.~\cite{Fleming:2007rp}, using an EFT, the results of 
Ref.~\cite{Voloshin:2003nt} was reproduced at LO, and the size of corrections to
the LO calculation was estimated. These
next-to-leading-order (NLO) corrections to the decay width include
effective-range corrections as well as calculable non-analytic
corrections from $\pi^0$ exchange. It was found that non-analytic
calculable corrections from pion exchange are negligible and the NLO
correction was dominated by contact interaction contributions. The
smallness of these corrections confirms one of the main points raised
in ~\cite{Fleming:2007rp}, namely, that pion exchange can be dealt
with using perturbation theory\footnote{This result has been also
  confirmed in Refs.~\cite{Baru:2011rs,Nieves:2012tt} and
  \cite{Valderrama:2012jv}. In the latter reference, the range of
  center-of-mass momenta for which the tensor piece of the one pion
  exchange potential is perturbative is studied in detail, and it is
  also argued that the effect of coupled channels is suppressed by at
  least two orders in the EFT expansion.}. However, the
$D\bar D$ FSI effects were not considered in these two works. 

The paper is structured as follows: in Section~\ref{sec:interaction}, we briefly 
discuss the $\X$ resonance within the hadronic molecular picture and the $S$-wave 
low-energy interaction between a charm and an anti-charm mesons. The decay 
$\X\to \DD\pi^0$ is discussed in detail in Section~\ref{sec:ddpi} with the 
inclusion of the $D\bar D$ FSI. Section~\ref{sec:summary} presents a brief 
summary.

\section{The $\bm{X(3872)}$ and the heavy meson $\bm S$-wave interaction}
\label{sec:interaction}

The basic assumption in this work is that the $\X$ exotic
charmonium is a $D\bar D^*-D^*\bar D$ bound state with quantum numbers
$J^{PC}=1^{++}$. For the sake of completeness we 
briefly discuss
in this section the formalism used in \cite{Nieves:2012tt,HidalgoDuque:2012pq}
to describe this resonance, which is based on solving and finding the poles of
the Lippmann-Schwinger equation (LSE). More specific details
can be found in these two references.

We use the matrix field $H^{(Q)}$ [$H^{(\bar Q)}$] to describe the
combined isospin doublet of pseudoscalar heavy-meson [antimeson]
$P^{(Q)}_a=(P^0,P^+)$ [$P^{(\bar Q)}_a=(\bar P^0,P^-)$] fields and
their vector HQSS partners $P^{*(Q)}_a$ [$P^{*(\bar Q)}_a$] (see for
example \cite{Falk:1990yz}),
\begin{eqnarray}
H_a^{(Q)} &=& \frac{1+\vsl}2 \left (P_{a\mu}^{* (Q)}\gamma^\mu -
P_a^{(Q)}\gamma_5 \right), \qquad v\cdot P_{a}^{* (Q)} = 0,  \nonumber \\
H^{(\bar Q)}_a &=&  \left (P_{a\mu}^{* (\bar Q)}\gamma^\mu -
P^{(\bar Q)}_a\gamma_5 \right) \frac{1-\vsl}2 , \qquad v\cdot P^{*
  (\bar Q)}_a = 0.
\end{eqnarray}
The matrix field $H^{c}$ [$H^{\bar c}$] annihilates $D$ [$\bar D$]
and $D^*$ [$\bar D^*$] mesons with a definite velocity $v$. The field
$H_a^{(Q)}$ [$H^{(\bar Q)}_a$] transforms as a $(2,\bar 2)$ [$(\bar
  2,2)$] under the heavy spin $\otimes $ SU(2)$_V$ isospin
symmetry~\cite{Grinstein:1992qt}. The definition for $H_a^{(\bar Q)}$ also 
specifies our convention for charge conjugation, which is $\mathcal{C}P_a^{(Q)} 
\mathcal{C}^{-1} = P^{(\bar Q) a} $ and $\mathcal{C}P_{a\mu}^{*(Q)}\mathcal{C}^{-1} 
= -P_\mu^{*(\bar Q) a} $. At very low energies, the
interaction between a charm and anti-charm meson can be accurately
described just in terms of a contact-range potential.
The LO Lagrangian respecting HQSS reads~\cite{AlFiky:2005jd}
\begin{eqnarray}
\label{eq:VLO}
\mathcal{L} & = & \frac{C_{A}}{4}\,\Tr\left[\bar{H}^{(Q)a}{H}_a^{(Q)} \gamma_{\mu} 
\right] \Tr\left[{H}^{(\bar{Q})a} \bar{H}^{(\bar{Q})}_a \gamma^{\mu} \right] 
\nonumber\\ &+& 
\frac{C_{A}^{\tau}}{4}\,\Tr\left[\bar{H}^{(Q)a} \vec\tau_a^{\,b} 
{H}^{(Q)}_{b} \gamma_{\mu} \right] \Tr\left[{H}^{(\bar{Q})c} 
\vec\tau_c^{\,d}\bar{H}^{(\bar{Q})}_{d} \gamma^{\mu} \right]  
\nonumber\\
&+& \frac{C_{B}}{4}\,\Tr\left[\bar{H}^{(Q)a}{H}_a^{(Q)} \gamma_{\mu}\gamma_5 
\right] \Tr\left[{H}^{(\bar{Q})a} \bar{H}^{(\bar{Q})}_a \gamma^{\mu}\gamma_5 
\right] \nonumber\\
&+&  
\frac{C_{B}^{\tau}}{4}\,\Tr\left[\bar{H}^{(Q)a} \vec\tau_a^{\,b} 
{H}^{(Q)}_{b} \gamma_{\mu}\gamma_5 \right] \Tr\left[{H}^{(\bar{Q})c} 
\vec\tau_c^{\,d}\bar{H}^{(\bar{Q})}_{d} \gamma^{\mu} \gamma_5\right]
\end{eqnarray}
with the hermitian conjugate fields defined as $\bar H^{ Q(\bar Q)
} =\gamma^0 H^{Q(\bar Q)\dag}\gamma^0$, and $\vec\tau$ the Pauli
  matrices in isospin space.  Note that in our normalization the heavy meson or
  antimeson fields, $H^{(Q)}$ or $H^{(\bar Q)}$, have dimensions of
  $E^{3/2}$ (see \cite{Manohar:2000dt} for details). This is because
  we use a non-relativistic normalization for the heavy mesons, which
  differs from the traditional relativistic one by a factor
  $\sqrt{M_H}$.
For later use, the four LECs that appear above  are rewritten into 
$C_{0A}$, $C_{0B}$ and $C_{1A}$, $C_{1B}$ which stand for the
counter-terms in the isospin $I=0$ and $I=1$ channels, respectively. The 
relations read
\begin{equation} 
C_{0\phi} = C_{\phi} + 3 C_{\phi}^{\tau}, \qquad
C_{1\phi} = C_{\phi} - C_{\phi}^{\tau}, \qquad \text{for}~ \phi = A,B
\end{equation}
The LO Lagrangian determines the contact interaction potential $V=i \mathcal{L}$, which 
is then used as kernel of the two body elastic LSE~\footnote{The
  extension to the general case of coupled channels is
  straightforward, as long as only two body channels are considered:
  $T$, $V$ and the two particle propagator will
  become matrices in the coupled channel space, being the latter one diagonal.} 
\begin{equation}
\label{eq:lse}
    T(E; \vec{p}\,',\vec{p}\,) = V(\vec{p}\,',\vec{p}\,) + \int 
\frac{d^3\vec{q}}{(2\pi)^3} V(\vec{p}\,',\vec{q}\,)
\frac{1}{E-\vec{q}^{\,2}/2\mu_{12}-M_1-M_2 +  i \epsilon} \, 
T(E; \vec{q},\vec{p}\,).
\end{equation}
with $M_1$ and $M_2$ the masses of the involved mesons, 
$\mu_{12}^{-1}=M^{-1}_1+M^{-1}_2$, $E$  the center of mass~(c.m.)
energy of the system and  $\vec{p}$ ($\vec{p}^{\, \prime}$)  the 
initial (final) c.m. momentum. Above threshold, we have $E> (M_1+M_2)$, 
and the unitarity relation ${\rm Im}\, T^{-1}(E) = \mu_{12}k/(2\pi)$ with $k= 
\sqrt{2\mu_{12}\left(E-M_1-M_2\right)}$.  
 
When contact interactions are used, the LSE shows an ill-defined
ultraviolet (UV) behaviour, and  requires a regularization and
renormalization procedure. We employ a standard Gaussian regulator
\begin{equation}
\left<\vec{p}\,|V|\vec{p}\,'\right> = C_{I\phi}
~e^{-\vec{p}\,^{2}/\Lambda^{2}}~e^{-\vec{p}\,'^{2}/\Lambda^{2}},
\end{equation}
with $C_{I\phi}$ the corresponding counter-term deduced from the
Lagrangian of Eq.~(\ref{eq:VLO}). We will take cutoff values $\Lambda =
0.5-1$ GeV~\cite{HidalgoDuque:2012pq,Nieves:2012tt}, where the range is chosen 
such that $\Lambda$ will be bigger than the
wave number of the states, but at the same time will be small enough
to preserve HQSS and prevent that the theory might become sensitive to
the specific details of short-distance dynamics. The dependence of
results on the cutoff, when it varies within this window, provides an
estimate of the expected size of subleading corrections.  On the other
hand, in the scheme of Ref.~\cite{HidalgoDuque:2012pq} pion exchange
and coupled-channel effects are not considered at LO. This is
justified since both effects were shown to be small by the explicit
calculation carried out in \cite{Nieves:2012tt} and the power counting
arguments established in \cite{Valderrama:2012jv}. Moreover, in what
pion exchange respects, this is in accordance with the findings of
Refs.~\cite{Fleming:2007rp, Baru:2011rs}, as mentioned in the
Introduction. 

Bound states correspond to poles of the $T$-matrix below threshold on the real 
axis in the first Riemann sheet of the complex energy. 
If we assume that the $\X$ state and the isovector $Z_b(10610)$ 
states~\footnote{The $Z_b(10610)$ observed in 
Ref.~\cite{Belle:2011aa} carries electric charge, and its neutral partner was 
also reported by the Belle Collaboration~\cite{Adachi:2012im}. We thus assume 
that its isospin is 1. } are $\left( D\bar D^*-D^* \bar D\right)/\sqrt{2}$ and 
$\left(B\bar B^*+B^* \bar B\right)/\sqrt{2}$ bound states, respectively, and 
use the isospin breaking information of the decays of the $\X$ into the
$J/\psi\pi\pi$ and $J/\psi\pi\pi\pi$, we can determine three linear
combinations among the four LECs $C_{0A}$, $C_{0B}$, $C_{1A}$ and 
$C_{1B}$ with the help of heavy quark spin and 
flavor symmetries~\cite{HidalgoDuque:2012pq, Guo:2013sya}.
We consider both the neutral $\left(D^0\bar D^{*0}-D^{*0} \bar D^0\right)$ and 
charged $\left(D^+D^{*-}-D^{*+} D^-\right)$ 
components in the $\X$. The coupled-channel potential is given by
\begin{eqnarray}
\label{eq:PotX}
V_{\X} = \frac{1}{2} \left( \begin{array}{cc}  C_{0X} + C_{1X} & C_{0X} - 
C_{1X} \\ C_{0X} - C_{1X} & C_{0X} + C_{1X}     \end{array} \right),
\end{eqnarray}
where $C_{0X}\equiv C_{0A}+C_{0B}$ and $C_{1X}\equiv C_{1A}+C_{1B}$.
Using $M_{\X}=(3871.68 \pm 0.17)$~MeV, the isospin violating
ratio of the decay amplitudes for the $\X\to J/\psi\pi\pi$ and $\X\to 
J/\psi\pi\pi\pi$, $R_{\X}=0.26\pm 0.07$~\cite{Hanhart:2011tn},\footnote{We 
have symmetrized the errors provided in \cite{Hanhart:2011tn} to use Gaussian 
distributions to estimate errors.} and the mass of the $Z_b(10610)$ (we assume 
that its
binding energy is $(2.0 \pm 2.0)$~MeV~\cite{Cleven:2011gp}) as 
three independent inputs, we find 
\begin{eqnarray}
 C_{0X}&=&-1.71^{+0.06}_{-0.04}~(-0.73^{+0.02}_{-0.01})\, {\rm fm}^{2}, 
\nonumber \\ 
C_{1X}&=&-0.13^{+0.53}_{-0.41}~(-0.39\pm0.09)\, {\rm 
fm}^{2}, \nonumber \\ 
C_{1Z}\equiv C_{1A}-C_{1B}&=&-0.75^{+0.24}_{-0.14}~(-0.30^{+0.03}_{-0.03})\, 
{\rm fm}^{2}
\label{eq:cvalues}
\end{eqnarray}
for $\Lambda = 0.5(1.0)$ GeV. Errors were obtained from a Monte Carlo (MC) 
simulation assuming uncorrelated Gaussian errors for the three inputs and using 
1000 samples.  Note that the values of the different LEC's are natural,
$\sim{\cal O}(1$~fm$^2)$, as one would expect. For details of the parameter 
determination, we refer to  
Refs.~\cite{HidalgoDuque:2012pq,Guo:2013sya,Guo:2013xga}. 

The $\X$ coupling constants to the neutral and charged 
channels, $g_0^X$ and $g_c^X$, respectively, are determined by the residues of 
the $T$-matrix elements at the $\X$ pole
\begin{eqnarray}
\label{eq:geq}
    \left(g_0^X\right)^2 &=& \lim_{E\to M_\X} \left[ E-M_\X \right] \times T_{11}(E),  \nonumber\\
    g_0^Xg_c^X &=& \lim_{E\to M_\X} \left[ E-M_\X \right] \times T_{12}(E),
\end{eqnarray}
where $T_{ij}$ are the matrix elements of the $T$-matrix solution of
the UV 
regularized  LSE. 
Their values are slightly different. Using the central values of $C_{0X}$ and 
$C_{1X}$, we get
\begin{eqnarray}
\label{eq:gval}
    g_0^X = 0.35^{+0.08}_{-0.29}\,(0.34^{+0.07}_{-0.29})~\text{GeV}^{-1/2},\qquad
g_c^X = 0.32^{+0.07}_{-0.26}\,(0.26^{+0.05}_{-0.22}) ~{\rm{GeV}}^{-1/2},
\end{eqnarray}
where, again, the values outside and inside the parentheses are
obtained with $\Lambda=0.5$ and 1~GeV, respectively. Note that when
the position of the $\X$ resonance approaches the $\DD$ threshold,
both couplings $g_0^X$ and $g_c^X$ vanish proportionally to the square
root of the binding energy~\cite{Gamermann:2009uq, Toki:2007ab}, which
explains the asymmetric errors.  Notice that the values of the 
coupling constants carry important information on the structure of the 
$X(3872)$. In general, the wave function of the $X(3872)$ is a composition of 
various Fock states, including the $c\bar c$, $D\bar D^*-D^*\bar D$, $c\bar c 
q\bar q~(q=u,d)$ and so on. The coupling constants are a measure of the 
probability of the $X(3872)$ to be a hadronic molecule~\cite{Gamermann:2009uq} 
(for discussions on the relation of the coupling constant with the composite 
nature of a physical state, we refer to 
Refs.~\cite{Weinberg:1965,Baru:2003qq,Hyodo:2013nka}).  

Within this model, we will account for the $D\bar D$ FSI effects to the $\X \to \DD\pi^0$ decay width. The $S$-wave interaction in the $D\bar D$ system with $J^{PC}=0^{++}$ 
is not entirely determined by $C_{0X}$, $C_{1X}$ and $C_{1Z}$. 
Indeed, considering again both the neutral and charged channels $\DD$ and 
$D^+D^-$, the potential is given by\footnote{The reason for using particle 
basis, where the interaction is not diagonal, instead of isospin basis is 
because for some values of the 
LEC's, a $\DD$ bound state close to threshold might be generated. If its binding energy is smaller or comparable to the $\DD-D^+D^-$ threshold difference, as it happens in the case of the $\X$ resonance, then 
it will become necessary to account for the mass difference among the neutral and charged channels.}
\begin{eqnarray}
\label{eq:vdd}
V_{D\bar{D}} = \frac{1}{2} \left( \begin{array}{cc}  C_{0A} + C_{1A} & C_{0A} - 
C_{1A} \\ C_{0A} - C_{1A} & C_{0A} + C_{1A}     \end{array} \right).
\end{eqnarray}
Thus, this interaction is not completely determined from what we have learned 
from the $\X$ and $Z_b(10610)$ states even if we use heavy quark spin and 
flavor symmetries --- the value of $C_{0A}$ is still unknown. Depending on the 
value of $C_{0A}$, there can be a $D\bar D$ $S$-wave bound state or not. For 
instance, considering the case for $\Lambda=0.5$~GeV and taking the central 
value for $C_{1A}=-0.44$~fm$^2$, if $C_{0A}=-3.53$~fm$^2$, then one finds a 
bound state pole in the $D\bar D$ system with a mass 3706~MeV (bound by around 25 MeV); if 
$C_{0A}=-1.65$~fm$^2$, there will be a $\DD$ bound state at threshold; if 
the value of $C_{0A}$ is larger, there will be no bound state pole any more. 
Therefore, the information of $C_{0A}$ will be crucial in understanding the 
$D\bar D$ system and other systems related to it through heavy quark 
symmetries~\cite{Guo:2013sya,Guo:2013xga}. Conversely, as we will see, the $\X \to \DD\pi^0$ decay width could be used
to extract information on the fourth LEC, $C_{0A}$, thanks to the FSI effects.

\section{ \bm{$\X\to \DD\pi^0$} decay}
\label{sec:ddpi}

\begin{figure}[tb]
\begin{center}
\includegraphics[width=\linewidth]{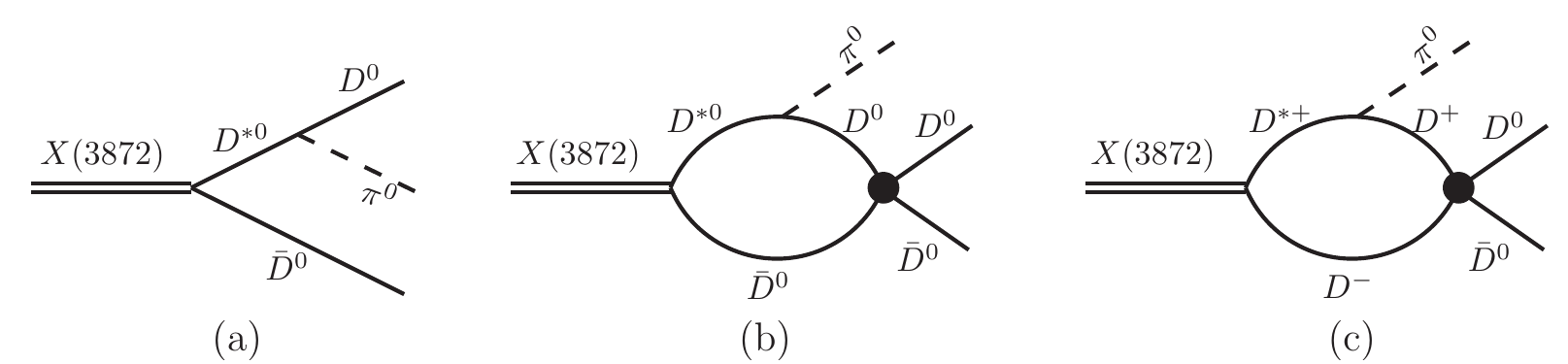}
\caption{Feynman diagrams for the decay $\X\to\DD\pi^0$. The charge conjugate
channel is not shown but included in the calculations.}\label{fig:FeynmanDiagrams}
\end{center}
\end{figure}

Here, we discuss the decay of the $\X$ into the  $\DD\pi^0$ mode. This decay can  
take place directly through the decay of the constituent $D^{*0}$ or $\bar 
D^{*0}$ as shown in Fig.~\ref{fig:FeynmanDiagrams}(a). After emitting a pion, 
the vector charm meson transit into a pseudoscalar one, and it can interact with 
the other constituent in the $\X$ 
as shown in Fig.~\ref{fig:FeynmanDiagrams}(b). 
Figure~\ref{fig:FeynmanDiagrams}(c) presents another possibility, namely the 
decay can also occur through the decay of the charged vector charm meson, 
and the virtual charged $D^+D^-$ pair then rescatter into $\DD$.

We will use the
relevant term in the LO Lagrangian of heavy meson chiral perturbation
theory~\cite{Grinstein:1992qt,Wise:1992hn,Burdman:1992gh,Yan:1992gz}
to describe the $D^{*} D \pi$ coupling
\begin{eqnarray}
{\cal L}_{\pi HH} &=& -\frac{g}{2 f_\pi} \left ( {\rm Tr} \left [\bar H^{(Q)b}
  H^{(Q)}_a \gamma_\mu \gamma_5\right] + {\rm Tr} \left [
  H^{(\bar Q)b}\bar H^{(\bar Q)}_a \gamma^\mu \gamma_5\right] \right)
(\vec\tau \partial_\mu  \vec\phi)^{\, a}_b  + \cdots \label{eq:LpiHH}
\end{eqnarray}
with $\vec\phi$ a relativistic field that describes the
pion\footnote{We use a convention such that $\phi= \frac{\phi_x-{\rm
      i} \phi_y}{\sqrt 2}$ creates a $\pi^-$ from the vacuum or
  annihilates a $\pi^+$, and the $\phi_z$ field creates or annihilates
  a $\pi^0$.}, $g\simeq 0.6$ is the $PP^*\pi$ coupling and $f_{\pi}
= 92.2\,{\rm MeV}$ the pion decay constant. Note that in our
normalization, the pion field has a dimension of energy, while the
heavy meson or antimeson fields $H^{(Q)}$ or $H^{(\bar Q)}$ have
dimensions of $E^{3/2}$, as we already mentioned.

\subsection{Tree Level Approximation}

For the process in question, the charm mesons are highly non-relativistic, thus we 
can safely neglect higher order terms in  $\vec{p}_{\bar 
D^{*0},D^{*0}}/M_{D^{*}}$.
Taking into account the contributions from both the $D^0\bar D^{*0}$ and 
$D^{*0}\bar D^0$ components of the $\X$, the tree-level amplitude is given by 
\begin{equation}
\label{eq:TreeLevel}
T_\text{tree} = - 2i \frac{gg_0^X}{f_{\pi}}
\sqrt{M_{X}} M_{D^{*0}} M_{D^0} \vec{\epsilon}_{X} \cdot 
\vec{p}_{\pi} \left( \frac{1}{p_{12}^{2} - M_{D^{*0}}^{2}} + 
\frac{1}{p_{13}^{2} - M_{D^{*0}}^{2}} \right),
\end{equation}
where $\vec\epsilon_X$ is the polarization vector of the $\X$, $\vec
p_\pi$ is the three-momentum of the pion, $p_{12}$ and $p_{13}$ are
the four momenta of the $\pi^0D^0$ and $\pi^0\bar D^0$ systems,
respectively\footnote{To obtain the amplitude, we have multiplied by
  factors $\sqrt{M_{D^{*0}}M_{D^0}}$ and $\sqrt{8 M_X
    M_{D^{*0}}M_{D^0}}$ to account for the normalization of the heavy
  meson fields and to use the coupling constant $g_0^X$, as defined in
  Eq.~\eqref{eq:geq} and given in Eq.~\eqref{eq:gval}, for the $\X
  D^0\bar D^{*0}$ and $\X D^{*0}\bar D^0$ vertices.}. We have
neglected the $D^{*0}$ and $\bar D^{*0}$ widths in the above
propagators because their inclusion only leads to small numerical
variations in the $\X \to \DD\pi^0$ decay rate of the order of 0.1
keV. As we will see below in Eq.~\eqref{eq:restree}, uncertainties on
the predicted width induced by the errors in the coupling $g_0^X$ and
the mass of the $\X$ resonance, turn out be much larger (of the order
of few keV).

Note that we have approximated the $\X D^0 \bar D^{*0}$ vertex by $g_0^X$. It 
could have some dependence on the momentum of the mesons, which can be
expanded in 
powers of momentum in the spirit of EFT. For the process in question , the 
momenta of the charm mesons are much smaller than the hard energy scale of the 
order of the cut-off, we can safely keep only the leading constant term.

Since the amplitude of Eq.~(\ref{eq:TreeLevel}) depends only on the
invariant masses $m_{12}^2 = p_{12}^2$ and $m_{23}^2=
(M^2_X+m^2_{\pi^0}+2M_{D^0}^2-m_{12}^2-p_{13}^2)$   of the final
$\pi^0D^0$ and $\DD$ pairs, respectively, we can use the standard form for the Dalitz plot~\cite {Beringer:1900zz}
\begin{equation}
d\Gamma = \frac{1}{(2\pi)^3}\frac{1}{32 M_X^3} \overline{|T|}^2
dm_{12}^2 dm_{23}^2
\end{equation}
and thus, we readily obtain
\begin{eqnarray}
\Gamma_\text{tree} &=& \frac{g^2}{192\pi^3 f_\pi^2}\left (g_0^X \frac{M_{D^0}
  M_{D^{*0}}}{M_X} \right)^2 \nonumber \\
&\times &\int^{(M_X-M_{D^0})^2}_{(M_{D^0}+m_\pi^0)^2} dm_{12}^2
\int^{(m^2_{23})_{\rm (max)}}_{(m^2_{23})_{\rm (min)}}d m^2_{23}\left( \frac{1}{p_{12}^{2} - M_{D^{*0}}^{2}} + 
\frac{1}{p_{13}^{2} - M_{D^{*0}}^{2}} \right)^2
|\vec{p}_\pi|^2 
\end{eqnarray}
with 
\begin{equation}
 |\vec{p}_\pi| = \frac{\lambda^{1/2}(M_X^2,m^2_{23},m_{\pi^0}^2)}{2M_X}
\end{equation}
the pion  momentum in the $\X$  center of mass frame [
$\lambda(x,y,z)=x^2+y^2+z^2-2(xy+yz+xz)$ is the K\"all\'en
function]. In addition, for a given value of $m^2_{12}$, the range of $m^2_{23}$ is determined by its
values when $\vec{p}_D$ is parallel or anti-parallel to $\vec{p}_{\bar D}$~\cite {Beringer:1900zz}:
\begin{eqnarray}
(m^2_{23})_{\rm (max)} & = & (E^*_D+E^*_{\bar D})^2 -(p^*_D-p^*_{\bar
    D})^2 \nonumber \\
(m^2_{23})_{\rm (min)} & = & (E^*_D+E^*_{\bar D})^2 -(p^*_D+p^*_{\bar
    D})^2  
\end{eqnarray}
with $E^*_D = (m_{12}^2-m^2_{\pi^0}+M_{D^0}^2) /2m_{12}$ and
$E^*_{\bar D} =
(M^2_X-m_{12}^2-M_{D^0}^2) /2m_{12}$ the energies
of the $D^0$ and $\bar D^0$ in the $m_{12}$ rest frame, respectively, and
$p^*_{D,\bar D}$ the moduli of their corresponding three momenta.

Using the couplings given in Eq.~\eqref{eq:gval},
the partial decay width for the three-body decay $\X\to 
D^0\bar D^0\pi^0$ at tree level is predicted to be
\begin{equation}
\label{eq:restree}
    \Gamma(\X\to D^0\bar D^0\pi^0)_\text{tree} = 
44.0_{-7.2}^{+2.4} 
\left( 42.0_{-7.3}^{+3.6} \right) ~\text{keV},
\end{equation}
where the values outside and inside the parentheses are obtained
with $\Lambda=0.5$ and 1~GeV, respectively, and the uncertainty
reflects the uncertainty in the inputs ($M_{\X}$ and the ratio
$R_{\X}$ of decay amplitudes for the $\X\to J/\psi\rho$ and $\X\to
J/\psi\omega$ decays). We have performed a Monte Carlo simulation to
propagate the errors.

Before studying the effects of the $D\bar{D}$ FSI, we would like to make two 
remarks:
\begin{enumerate}
 \item Within the molecular wave-function description of the $\X \to
   \DD\pi^0$ decay, the amplitude of Fig.~\ref{fig:FeynmanDiagrams}(a) 
will read\footnote{For simplicity, we omit the contribution to the amplitude driven 
by the ${D}^{*0}\bar D^0 $ component of the $\X$ resonance, for which the discussion will run in parallel. } 
\begin{eqnarray}
 T_\text{tree} &\sim& \int d^3 q \underbrace{\langle D^0\bar{D}^{*0} 
(\vec{p}_{D^0} )| D^0\bar{D}^{*0} (\vec{q}\,)\rangle}_{\propto 
\delta^3\left({\vec{q}-\vec{p}_{D^0}}\right)} 
\langle  D^0\bar{D}^{*0}(\vec{q}\,) | \X \rangle T_{{\bar 
D}^{0*}(\vec{p}_{\bar{D}^{0*}})\to \bar D^0 \pi^0} \nonumber\\
&= &
\Psi(\vec{p}_{D^0})T_{{\bar D}^{0*}(\vec{p}_{\bar{D}^{0*}})\to \bar D^0 \pi^0} 
\label{eq:tree.wf}
\end{eqnarray}
with $\vec{p}_{\bar{D}^{0*}}=-\vec{p}_{D^0}$ in the laboratory frame. Note that this description is totally equivalent to that of Eq.~(\ref{eq:TreeLevel}) because the $D^0\bar D^{*0}$ component of the non-relativistic 
$\X$ wave-function 
is given by~\cite{Gamermann:2009uq} 
\begin{equation}
 \Psi(\vec{p}_{D^0}) = \frac{g_0^X}{E_{D^0}+E_{\bar D^{*0}}-M_{D^0}-M_{ 
D^{*0}}- \vec{p}_{D^0}^{\,2}/2\mu_{D^0D^{*0}}} = \frac{g_0^X}{E_{\bar 
D^{*0}}-M_{ D^{*0}}- \vec{p}_{\bar D^{*0}}^{\,2}/2M_{ D^{*0}}}
\end{equation}
with $\mu^{-1}_{D^0D^{*0}}= M_{D^0}^{-1}+M_{ D^{*0}}^{-1}$. In the last step, we have used that the $D^0$ meson is on shell and therefore $ \left(E_{D^{0}}-M_{ D^{0}}- \vec{p}_{ D^{0}}^{\,2}/2M_{ D^{0}}\right) =0$. 
Thus, the wave function in momentum space  turns out to be proportional to the coupling  $g_0^X$ times the non-relativistic reduction, up to a factor $2M_{ D^{*0}}$, of the $\bar D^{* 0}$ propagator 
that appears in Eq.~(\ref{eq:TreeLevel}).

The amplitude in Eq.~(\ref{eq:tree.wf}) involves the $\X$ wave function at a given momentum, $\vec{p}_{D^0}$, and the total decay width depends on the wave function in momentum space evaluated only for a limited range of values of 
$\vec{p}_{D^0}$  determined by energy-momentum conservation.  This is in sharp 
contrast to the decay amplitude into charmonium states, as shown in 
Fig.~\ref{fig:shortdis}, in Eq.~(\ref{eq:psi0}),  where there is an integral 
over all possible momenta included in the wave function. Such an integral can 
be thought of as a Fourier transform at $\vec{x}=0$, and thus gives rise to the 
$\X$ wave function in coordinate space at the origin. This is to say, the width 
is proportional to the probability of finding the $D\bar D^*$ pair at zero 
(small in general) relative distance within the molecular $\X$ state. This 
result is intuitive, since the $D\bar D^*$ transitions to final states 
involving charmonium mesons should involve the exchange of a virtual charm 
quark, which is only effective at short distances. However, in the  $\X \to 
\DD\pi^0$ process, the relative
distance of the $D\bar D^*$ pair can be as large as
allowed by the size of the $\X$ resonance, since the final state is
produced by the one body decay of the $\bar D^*$ meson instead of by a 
strong two body transition. Thus, this decay channel might provide 
details on the long-distance part of the $\X$ wave function. Indeed, from 
Eq.~(\ref{eq:tree.wf}) it follows that a future measurement of the 
$d\Gamma/d|\vec{p}_{D^0}|$ distribution 
might provide valuable information on the $\X$ wave-function $\Psi(\vec{p}_{D^0})$.

\item So far, we have not made any reference to the isospin nature of the $\X$ resonance. We have just used the coupling,  $g_0^X$, of the resonance to the $D^0\bar D^{0*}$ pair. 
In addition  to the $J/\Psi\pi^+\pi^-\pi^0$   final state, the $\X$ decay into $J/\Psi\pi^+\pi^-$ was also observed \cite{Abe:2005ix,delAmoSanchez:2010jr}, pointing out to 
an isospin violation, at least,  in its decays~\cite{Tornqvist:2004qy}.  In the $D\bar D^*$ molecular picture, 
the isospin breaking effects arise due to the mass difference between the 
$D^0\bar D^{*0}$ pair and its charged counterpart, the $D^+\bar D^{*-}$ pair, 
which turns out to be relevant because of the closeness of the $\X$ mass to the $D^0D^{*0}$ threshold~\cite{Tornqvist:2004qy,Gamermann:2009fv,Gamermann:2009uq,HidalgoDuque:2012pq}.
The observed isospin violation in the decays $\X \rightarrow \rho J/\psi$, and 
$\X\rightarrow \omega J/\psi$  depends on the probability amplitudes of both 
the neutral and charged meson channels near the origin 
which are very similar~\cite{Gamermann:2009uq}. This suggests that, when dealing with these strong processes, the isospin $I=0$ component will be the most relevant, though the experimental value of the isospin 
violating ratio, $R_{\X}$, of decay amplitudes  could be used to learn details on the weak $D\bar D^*$ interaction in the isovector channel~\cite{HidalgoDuque:2012pq} ($C_{1X}$ in Eq.~(\ref{eq:PotX})). 
The $\X \to \DD\pi^0$ decay mode can shed more light 
into the isospin dynamics of the $\X$ resonance, since it can be used to further 
constrain the isovector sector of the $D\bar D^*$ interaction. This is the case 
already at tree level because the numerical value of the coupling 
$g_0^X$ is affected by the interaction in the isospin one channel, $C_{1X}$. 

We should also stress that in absence of FSI effects that will be discussed below, if $C_{1X}$ is neglected, as in Ref.~\cite{Gamermann:2009uq}, the $\X \to \DD\pi^0$ width will be practically the same 
independent of whether the $\X$ is considered as an isoscalar molecule or a 
$D^0\bar D^{*0}$ state. In the latter case, the width would be proportional to 
$\tilde g^2$ \cite{Gamermann:2009uq}, 
\begin{equation}
 \tilde g^2 = - \left. 
\left(\frac{dG_{0}}{dE}\right)^{-1}\right|_{E=M_X}, \qquad G_{0}(E) = 
\int_\Lambda 
 \frac{d^3\vec{q}}{(2\pi)^3} 
\frac{1}{E-M_{D^0}-M_{D^{*0}}-\vec{q}^{\,2}/2\mu_{D^0D^{*0}}} 
\end{equation}
where $G_0(E)$ is the UV regularized $D^0\bar D^{*0}$ loop function\footnote{Notice that 
although the loop function is linearly divergent, its derivative with
respect $E$ is convergent, and thus it only shows a  residual (smooth) dependence
on $\gamma/\Lambda$ if a gaussian cutoff is used, with 
$\gamma^2=2\mu_{D^0D^{*0}}(M_{D^0}+M_{D^{*0}}-M_X)$. Were a sharp cutoff 
used, there would be no any dependence on the cutoff because of the 
derivative.}. However,  if the $\X$ were an isoscalar state, 
\begin{equation}
 | \X \rangle = \frac{1}{\sqrt{2}}\left (|D^0\bar D^{*0}\rangle +  
|D^+D^{*-}\rangle\right) 
\end{equation}
one, naively, would expect to obtain a width around a factor two smaller, 
because now the coupling of the $\X$ state to the $D^0\bar D^{*0}$ pair would be 
around a factor $\sqrt 2$ smaller as well \cite{Gamermann:2009uq} 
\begin{equation}
 \left(g_{0}^X \right)^2 \simeq \left(g_{c}^X \right)^2 \simeq 
- \left. \left(\frac{dG_{0}}{dE}+ 
\frac{dG_{c}}{dE}\right)^{-1} \right|_{E=M_X}, 
\end{equation}
where $G_c$ is the loop function in the charged charm meson channel.
The approximations would become equalities if the isovector 
interaction is neglected (it is much smaller than the isoscalar one as can be 
seen in Eq.~\eqref{eq:cvalues}). 
Were $\frac{dG_{0}}{dE}\simeq\frac{dG_{c}}{dE} $, the above values would be 
equal to $\tilde g^2/2$ approximately. However, after considering the mass 
differences between the neutral and charged channels and, since 
$\frac{dG_{i}}{dE} \propto 1/\sqrt{B_i}$ [$B_i>0$ is the 
binding energy of either the neutral ($\sim 0.2$ MeV) or charged ($\sim 8$ MeV) 
channels], at the mass of the $\X$ one actually finds 
\begin{equation}
 \left. \left(\frac{dG_{0}}{dE}\right)\right|_{E=M_X} \gg \left. \left(\frac{dG_{c}}{dE}\right)\right|_{E=M_X} 
\end{equation}
so that $\left(g_{0}^X \right)^2 \simeq \tilde g^2$. Therefore, the prediction 
for the decay width would hardly change.

All these considerations are affected by the $D\bar D$ FSI effects which will be discussed next.

\end{enumerate}

\subsection {\bm{$D\bar{D}$} FSI Effects} 

To account for the FSI effects, we include in the analysis the $D
\bar{D}\to D\bar D$ $T$-matrix, which is obtained by solving the LSE
(Eq.~\eqref{eq:lse}) in coupled channels with the $V_{D\bar D}$
potential given in Eq.~\eqref{eq:vdd}. We use in Eq.~\eqref{eq:lse}
the physical masses of the neutral ($\DD$) and charged ($D^+D^-$)
channels. Thus, considering both the $D^0\bar D^{*0}$ and $D^{*0}\bar
D^0$ meson pairs as intermediate states, the decay amplitude for the
mechanism depicted in Fig.~\ref{fig:FeynmanDiagrams}(b) reads
\begin{equation}
\label{eq:loop0}
T_\text{loop}^{(0)} = - 16i \frac{gg_{0}^X}{f_{\pi}}  
\sqrt{M_{X}} M_{D^{*0}} M_{D^0}^{3} \,\vec{\epsilon}_{X} \cdot 
\vec{p_{\pi}} \, T_{00\to 00}(m_{23})\,
I(M_{D^{*0}},M_{D^{0}},M_{D^{0}},\vec{p}_\pi),
\end{equation}
where $T_{00\to 00}$ is the $T$-matrix element for the
$D^{0}\bar{D}^{0}\to D^{0}\bar{D}^{0}$ process, and the three-point loop function is 
defined as
\begin{eqnarray}
\label{3pointEq}
I\left(M_{1},M_{2},M_{3},\vec{p}_\pi\right) = i \int{ 
\frac{d^{4}q}{\left(2\pi\right)^{4} 
}   \frac{1}{q^{2}-M_{1}^{2}+i\varepsilon} \frac{1}{\left(P - 
q\right)^{2}-M_{2}^{2}+i\varepsilon}  
\frac{1}{\left(q - p_{\pi}\right)^{2}-M_{3}^{2}+i\varepsilon}},
\end{eqnarray}
with $P^\mu=(M_X,\vec{0})$ in the rest frame of the $\X$. This loop
integral is convergent. Since all the intermediate mesons in the
present case are highly non-relativistic, the three point loop can be
treated non-relativistically. The analytic expression for this loop
function at the leading order of the non-relativistic expansion can be
found in Eq.~(A2) of Ref.~\cite{Guo:2010ak} (see also
Ref.~\cite{Cleven:2011gp}). For the specific kinematics of this decay,
the loop function in the neutral channel  has an imaginary part, which
turns out to be much larger  than the real one,
except in a narrow region involving the highest pion momenta.

Similarly, the amplitude for the 
mechanism with charged intermediate charm mesons is given by
\begin{equation}
\label{eq:loopc}
T_\text{loop}^{(c)} = 16i \frac{gg_{c}^X}{f_{\pi}}  
\sqrt{M_{X}} M_{D^{*0}} M_{D^0} M_{D^\pm}^2 \,\vec{\epsilon}_{X} \cdot 
\vec{p_{\pi}} \, T_{+-\to 00}(m_{23})\,
I(M_{D^{*\pm}},M_{D^{\pm}},M_{D^{\pm}},\vec{p}_\pi),
\end{equation}
where  $T_{+-\to 00}$ is the $T$-matrix element for the $D^+D^-\to D^{0}\bar{D}^{0}$ 
process. The loop function is now purely real because the
$D^+D^-\pi^0$ channel is closed, and its size  is
significantly smaller than in the case of the neutral channel. The sign difference between the  amplitudes of Eqs.~(\ref{eq:loop0}) 
and ~(\ref{eq:loopc}) is due to the sign difference between the $D^{*-}  \to 
D^- \pi^0 $ and $\bar D^{*0}  \to D^0 \pi^0 $ transition amplitudes.
%

\begin{figure}[tb]
\centering
    \includegraphics[width=0.49\linewidth]{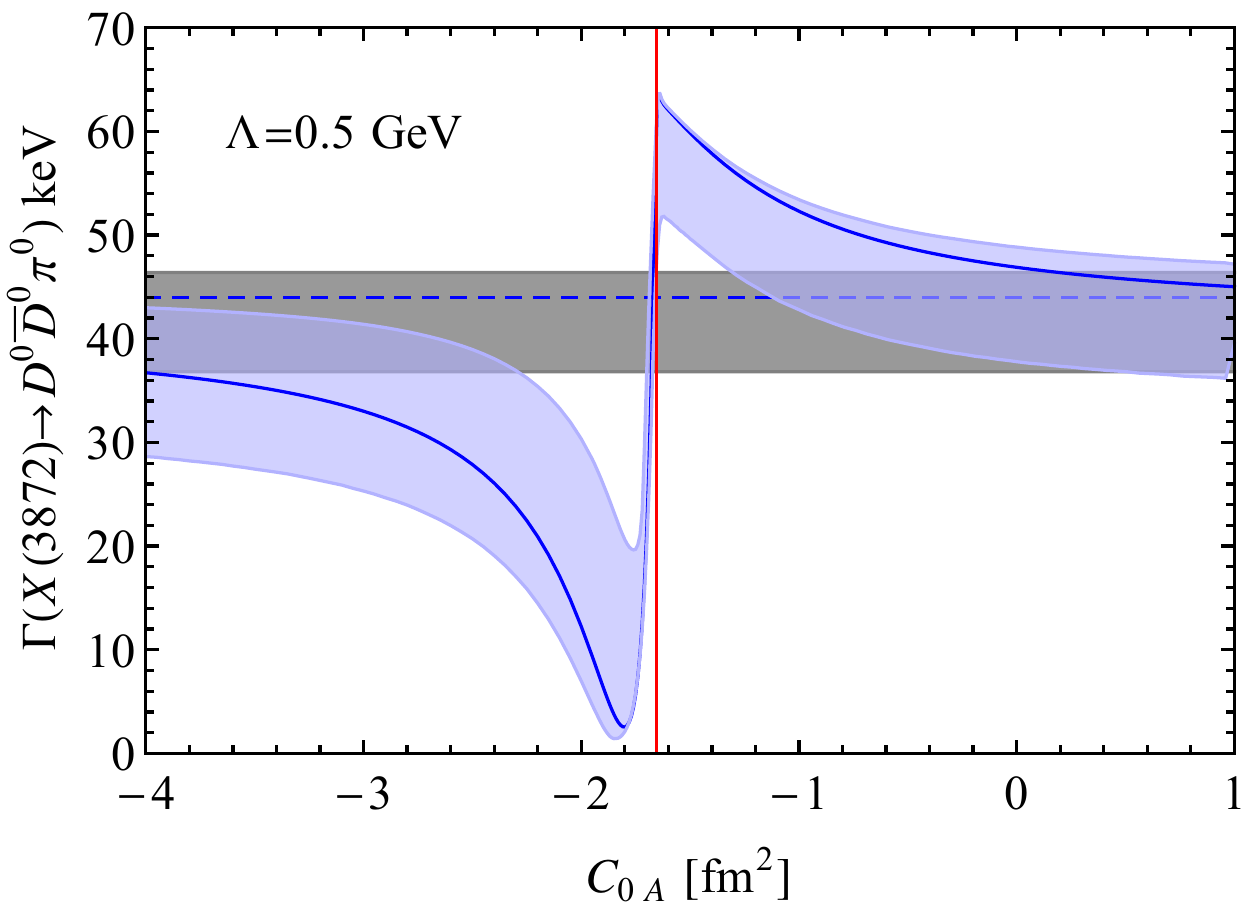}\hfill
    \includegraphics[width=0.49\linewidth]{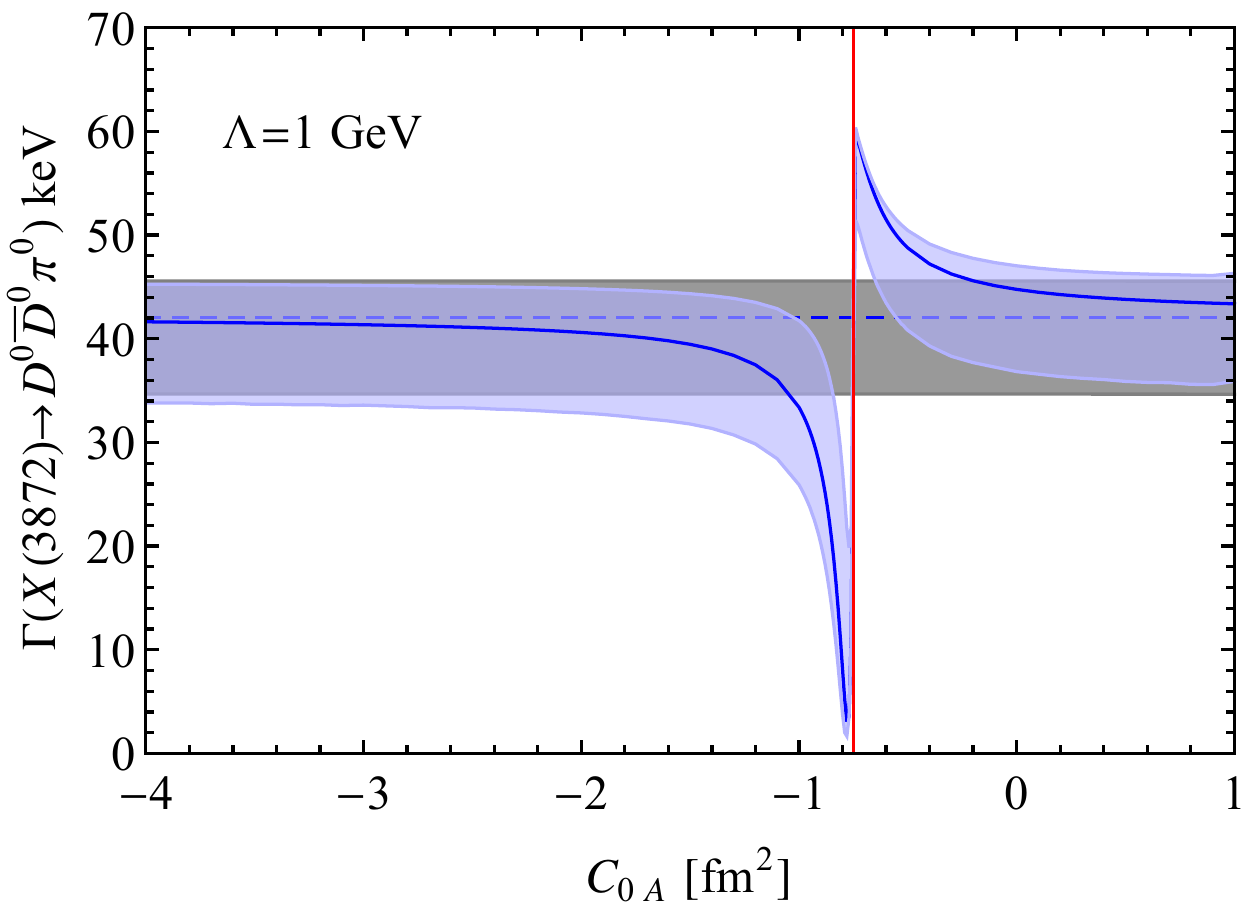}
    \caption{Dependence of the $\X\to D^0\bar D^0\pi^0$ partial decay
      width on the low-energy constant $C_{0A}$. The UV cutoff is set
      to $\Lambda=0.5$ GeV (1 GeV) in the left (right) panel.  The
      blue error bands contain $D\bar D$ FSI effects, while the grey
      bands stand for the tree level predictions of
      Eq.~(\ref{eq:restree}). The solid (full calculation) and dashed
      (tree level) lines stand for the results obtained with the
      central values of the parameters. The vertical lines denote the
      values of $C_{0A}$ for which a $D\bar D$ bound state is
      generated at the $D^0\bar D^0$ threshold.}\label{fig:c0a}
\end{figure}

For consistency, despite the three-point loop functions in
Eqs.~(\ref{eq:loop0})--(\ref{eq:loopc}) being finite, they should
however be evaluated using the same UV renormalization scheme as that
employed in the $D^{(*)} \bar D^{(*)}$ EFT. The applicability of the
EFT relies on the fact that long range physics should not depend on
the short range details. Hence, if the bulk of contributions of the
loop integrals came mostly from large momenta (above 1 GeV for
instance), the calculation would not be significant. Fortunately, this is not 
the case, and the momenta involved in the integrals are rather low. Indeed, 
the biggest FSI contribution comes from the imaginary
part of the loop function in the neutral channel, which is hardly
sensitive to the UV cutoff. Thus and for the sake of simplicity, FSI
effects have been calculated using the analytical expressions for the
three-point loop integral mentioned above, valid in the $\Lambda\to
\infty$ limit. Nevertheless, we have  numerically computed  these
loop functions with 0.5 and 1 GeV UV Gaussian cut-offs and found small
differences\footnote{The largest changes affect to the charged channel
  (Fig.~\ref{fig:FeynmanDiagrams}(c)). This is because there, the
  three meson loop integral is purely real. However this FSI
  mechanism, as we will discuss below, provides a very small
  contribution to the total decay width.  } in the final results
[$\Gamma(\X\to\DD\pi^0)$ versus $C_{0A}$] discussed in
Fig.~\ref{fig:c0a}. Indeed, the changes turn out to be almost
inappreciable for $\Lambda=1$ GeV, and they are at most of the order
of few percent in the $\Lambda=0.5$ GeV case. Moreover, even then, these differences are well
accounted for the error bands shown in the figure.

To compute $T_{00\to 00}$ and $T_{+-\to 00}$ we need the $D\bar D$
potential given in Eq.~\eqref{eq:vdd}. With the inputs (masses of the
$\X$ and $Z_b(10610)$ resonances and the ratio $R_{\X}$) discussed in
Section~\ref{sec:interaction}, three of the four couplings, that
describe the heavy meson-antimeson $S$-wave interaction at LO in the
heavy quark expansion, can be fixed. The value of the contact term
parameter $C_{0A}$ is undetermined, and thus we could not predict the
$D\bar D$ FSI effects parameter-free in this $\X$ decay.  These
effects might be quite large, because for a certain range of $C_{0A}$
values, a near-threshold isoscalar pole could be dynamically generated
in the $D\bar D$ system \cite{HidalgoDuque:2012pq, Nieves:2012tt}.

To investigate the 
impact of the  FSI, in Fig.~\ref{fig:c0a} we show the dependence of the partial 
decay width on $C_{0A}$. For comparison, the tree-level results 
are also shown in the same plots.  The vertical lines denote the 
values of $C_{0A}$ when there is a $D\bar D$ bound state at threshold. 
When $C_{0A}$ takes smaller values, the binding energy becomes larger; when 
$C_{0A}$ takes larger values, the pole moves to the second Riemann sheet and 
becomes a virtual state. Around the values denoted by the vertical lines, 
the pole is close to threshold no matter on which Riemann sheet it is. One can 
see an apparent deviation from the tree-level results in this region. The wavy 
behavior is due to the interference between the FSI and the tree-level terms.
The existence of a low lying $D\bar D$ bound state has as a consequence
a decrease of the partial decay width to $D^0 \bar{D}^0 \pi^0$,
the reason being that there's a substantial probability of
a direct decay to the $D\bar D$ bound state and a neutral pion.
On the other hand if there is a virtual state near the threshold,
the decay width will increase owing to rescattering effects~\footnote{
The mechanism is analogous for instance to the large capture cross-section
of thermal neutrons by protons or the near threshold enhancement of
deuteron photo-disintegration, both of which are triggered by
the existence of a virtual state in the singlet neutron-proton channel.}.

When  the partial decay width will be measured in future experiments, a significant 
deviation from the values in Eq.~\eqref{eq:restree} will indicate a FSI 
effect, which could eventually be used to extract the value of $C_{0A}$.  
Outside the wavy region, the FSI contribution is small, and it will be
unlikely to obtain any conclusive information on $C_{0A}$ from the experimental $\Gamma(\X\to\DD\pi^0)$ width. However, there exist theoretical hints
pointing out the existence of a $D \bar D$ bound state close to threshold.
In the scheme of Ref.~\cite{HidalgoDuque:2012pq}, the $Z_b(10610)$ mass input was not used, but however there, it was assumed that 
the $X(3915)$ and $Y(4140)$ were $D^*\bar{D}^*$and $D_s^*\bar{D}_s^*$ molecular states. These two new inputs were used to fix completely the 
heavy meson-antimeson interaction, and a 
$D \bar D$ molecular isoscalar state was predicted at around 3710~MeV. A state 
in the vicinity of
$3700\,{\rm MeV}$ was also predicted in Ref.~\cite{Gamermann:2006nm}, within 
the hidden gauge formalism,
using an extension of the SU(3) chiral Lagrangians to SU(4) that
implements a particular pattern of SU(4) flavor symmetry breaking. Experimentally, there is 
support for that resonance around 3720 MeV from the analysis of the $e^+ e^- \to J/\psi D \bar D$ Belle data~\cite{Abe:2007sya} carried out in  
\cite{Gamermann:2007mu}. However, the broad bump observed   above 
the $D \bar D$ threshold  by the Belle Collaboration in the previous 
reaction  could instead 
 be produced by the  $\chi_0(2P)$ state~\cite{Chao:2007it, Guo:2012tv}.

In Ref.~\cite{Aceti:2012cb}, the authors show that the charged component 
$D^+D^{*-}$ in the $\X$ is essential to obtain a width for the $\X\to 
J/\psi\gamma$ compatible with the data. In the process studied in this work, 
at tree-level, the charged component does not directly contribute, though it could  
indirectly modify the  $\X D^0\bar D^{(*0)}$  coupling $g_0^X$. However,
because the $\X$ resonance is placed so close to the $D^0\bar D^{(*0)}$ 
threshold,  we argued that this is not really the case and such a component 
hardly changes the prediction for the decay width. When the FSI is taken 
into account, one may ask whether the charged component is important or not 
since it can now contribute as the intermediate state which radiates the pion. 
We find, however, this contribution plays a small role here, leading
to changes of about  ten percent at most for the $\Lambda=0.5 $ GeV
case, and much smaller when the UV cutoff is set to 1
GeV. These variations are significantly smaller that the uncertainty
bands displayed in Fig.~\ref{fig:c0a}.  Therefore, we conclude that the relative importance 
of the charge component in the $\X$ depends on the process in question. When 
the observable 
is governed by the wave function of the $\X$ at the origin, it can be important 
as the case studied in Ref.~\cite{Aceti:2012cb}. For our case, the decay is 
more sensitive to the long-distance structure of the $\X$, then the charged 
component is not as important as the neutral one. At this point, we can also comment 
on the processes $\X\to \DD\gamma$ and $\X\to D^+D^-\gamma$, where the $D\bar 
D$ system has now a negative $C$ parity in contrast to the pionic decay. The 
decay 
amplitudes, when neglecting possible contributions from the $\psi(3770)$, are
similar to the one in Eq.~\eqref{eq:TreeLevel}. Near the $D\bar D$ threshold, 
the intermediate $D^{*0}$ is almost on shell, and the virtuality of the charged 
$D^{*}$ is much larger. Thus, the partial decay width into the $D^+D^-\gamma$ 
should be much smaller than the one into the $\DD\gamma$, as discussed in 
Ref.~\cite{Voloshin:2005rt}.

\section{Summary}
\label{sec:summary}

In this work, we explored the decay of the $\X$ into the $\DD\pi^0$ using 
an effective field theory based on the hadronic molecule assumption for the $\X$. 
This decay is unique in the sense that it is sensitive to the long-distance 
structure of the $\X$ as well as the strength of the $S$-wave interaction 
between the $D$ and $\bar D$. We show that if there was a near threshold pole in 
the $D\bar D$ system, the partial decay width can be very different from the 
result neglecting the FSI effects. Thus, this decay may be used to measure the 
so far unknown parameter $C_{0A}$ in this situation. Such information is 
valuable to better understand the interaction between a heavy and an anti-heavy 
meson. In view that some of the $XYZ$ states which are attracting intensive 
interests are good candidates for the heavy meson hadronic molecules, it is 
desirable to carry out a precise measurement of that width.

It is also worth mentioning that since this decay is sensitive to the 
long-distance structure, the 
contribution of the $\X $ charged component $\left(D^+D^{*-}-D^{*+}D^-\right)$ 
is not important even 
when the $D\bar D$ FSI is taken into account. We have also discussed how a future 
measurement of the $d\Gamma/d|\vec{p}_{D^0}|$ distribution 
might provide valuable information on the $\X$ wave function at the fixed 
momentum $\Psi(\vec{p}_{D^0})$.

\medskip

\section*{Acknowledgments}
We thank U.-G.~Mei{\ss}ner and E.~Oset for  useful discussions and for
a careful reading of the manuscript. C.~H.-D. thanks the support of the 
JAE-CSIC Program. F.-K.~G. would like to 
thank the Valencia group for their hospitality during his visit when part of 
this work was done. This work is supported in 
part by the DFG and the NSFC through funds provided to the Sino-German CRC 110 
``Symmetries and the Emergence of Structure in QCD'', by the NSFC (Grant No. 
11165005), by the Spanish Ministerio de Econom\'\i a y Competitividad and 
European FEDER funds under the contract FIS2011-28853-C02-02 and the Spanish 
Consolider-Ingenio 2010 Programme CPAN (CSD2007-00042), by Generalitat 
Valenciana under contract PROMETEO/2009/0090 and by the EU HadronPhysics3 
project, grant agreement no. 283286.

\end{document}